\documentclass[11pt]{article}
\usepackage[margin=1in]{geometry}                % See geometry.pdf to learn the layout options. There are lots.
\usepackage{subfigure,epsfig,amsfonts}
\usepackage{amsmath}
\usepackage{amssymb}
\usepackage{amsthm}
\usepackage{dsfont}
\usepackage[font=small, labelfont=bf]{caption}

\title{Turning Statistical Physics Models Into Materials Design Engines}
\author
       {Marc Z. Miskin$^1$, Gurdaman S. Khaira$^2$, Juan J. de Pablo$^2$, Heinrich M. Jaeger$^{1}$
              \\ \\
       $^1$James Franck Institute, The University of Chicago, \\929 East 57th St., Chicago, IL 60637, USA.\\
       $^2$Institute for Molecular Engineering, The University of Chicago, \\5801 South Ellis Ave, Chicago, IL 60637, USA.\\ \\
       }

\begin{document}
\maketitle

\begin{abstract}
Despite the success statistical physics has enjoyed at predicting the properties of materials for given parameters, the inverse problem, identifying which material parameters produce given, desired properties, is only beginning to be addressed.  Recently, several methods have emerged across disciplines that draw upon optimization and simulation to create computer programs that tailor material responses to specified behaviors.  However, so far the methods developed either involve black-box techniques, in which the optimizer operates without explicit knowledge of the material's configuration space, or they require carefully tuned algorithms with applicability limited to a narrow subclass of materials.  Here we introduce a formalism that can generate optimizers automatically by extending statistical mechanics into the realm of design. The strength of this new approach lies in its capability to transform statistical models that describe materials into optimizers to tailor them. By comparing against standard black-box optimization methods, we demonstrate how optimizers generated by this formalism can be faster and more effective, while remaining straightforward to implement. The scope of our approach includes new possibilities for solving a variety of complex optimization and design problems concerning materials both in and out of equilibrium.

\end{abstract}

\section{Significance Statement} {A fundamental tenet of science is that the properties of a material are intimately linked to the nature of the constituent components. While there are powerful methods to predict such properties for given components, a key challenge for materials design is the inverse process: identifying the required components and their structural configuration for given target properties. This paper presents a new approach to this challenge. A formalism is introduced that generates algorithms for materials design both under equilibrium and non-equilibrium conditions, and operates without the need for user input beyond a design goal.Ê This formalism is broadly applicable, fast and robust, and it provides a powerful new tool for materials optimization as well as discovery.Ê}
\section{Introduction}
Computer programs that can design material properties have lead to exciting, new directions for materials science\cite{jain2014inverse,torquato2009inverse,jaeger2015celebrating}.  Computational methods have been used to predict crystal\cite{oganov2006crystal,Oganov2011,oganov2008evolutionary}  and protein\cite{dahiyat1997novo,kuhlman2003design} structures, yielding the toughest crystals known to man \cite{Oganov2011} and de-novo protein configurations unseen in nature \cite{dahiyat1997novo}. Applied to polymers, Monte-Carlo methods \cite{hannon2013inverse,hannon2013optimizing,chang2014design} and evolutionary algorithms \cite{Qin2013,khaira2014evolutionary} have paved the way towards optimizing directed self-assembly.  Similar methods have been employed to identify the crystal structures of patchy, colloidal particles \cite{bianchicrystals}.  For far-from-equilibrium systems like jammed, metastable aggregates of particles  \cite{jaeger2015celebrating}, simulation-based optimization has been successfully used to design bulk properties like stiffness \cite{Miskin2013} and packing density \cite{miskin2014evolving}  by way of tuning complicated micro-scale features like particle shape.

Yet in spite of these successes, most of the existing methods work only for narrowly defined classes of materials: optimization techniques that prove successful at designing one class of materials may struggle or fail on other systems. Thus, designing new materials can require a large investment in trial and error at the level of the algorithm itself, even if, for given parameters, the material's behavior can be simulated easily.

In standard black-box approaches to optimization, the algorithm tunes the material by adjusting a set of control parameters without considering the likelihood of finding the material in micro-scale configurations.  Instead, the optimizer operates in some auxiliary space, defined outside the physical model, and remains ignorant of the statistics in the physical configuration space. On the other hand, for the overwhelming majority of materials, an accurate description of macro-scale behavior comes about by explicitly considering the probability of finding the system in certain microscopic configurations by way of a statistical mechanics model. Several materials optimization approaches exist that take this statistical nature into account, and examples include optimizers that design spin configurations\cite{PhysRevB.88.134104,solis2010controlled}, patchy colloidal particles\cite{jankowski2012}, and self-assembly driven by short-ranged interactions\cite{hormoz2011design}.  These approaches are carefully tailored and build heavily upon the specific model defining the material.  As a consequence, it is difficult to extend such approaches beyond the material class they are designed to work on. Evidently, micro-scale configurations present key statistical information about a material, which is completely ignored by black-box approaches, yet there is no formalism that generically incorporates this information into materials design.

The question we ask is whether micro-state information can be used not only to enhance an optimizer's speed and range of applicability, but also whether it can become the cornerstone of an approach that automatically transforms a design goal and a statistical model that accurately describes the likelihood of micro-states into an optimizer. One can then construct optimizers that can work on material classes that are as broad as those described by statistical mechanics, without the need for ad-hoc modification. In other words, this avoids the need to experiment with combinations of optimizers, tweak quality functions, or introduce theoretical simplifications.

Here we take some first steps towards such a framework.  We present a formalism that can be used to transform the capacity to predict material behavior into an optimizer that tunes it.  Furthermore we find that our formalism often solves optimization problems faster and more reliably than approaches built around black-box optimization methods.

\section{Deriving the Optimization Equations}

Our approach starts by assuming that we are given a model $\rho(x | \lambda_i)$ which predicts the probability of finding a material system in some configuration, $x$.  The model depends on the adjustable parameters $\lambda_i$ and by tuning $\lambda_i$, a user can impact the emergent, bulk properties, averaged over configuration space.  Thus, design proceeds by tweaking the values of each $\lambda_i$ to promote a desired, user-specified response.

We work towards this goal starting from the heuristic equation
\begin{equation}\label{rep1}
\dot{\rho}{(x|\lambda_i)}= \rho(x|\lambda_i)[f(x)-\langle f(x) \rangle]
\end{equation}
where the angle brackets denote an average over configurations weighted by $\rho$,   the overdot denotes a derivative with respect to an artificial time that indexes the optimization steps, and $f(x)$ is a function that quantifies how well configuration $x$ represents the user specified goal.   For example, if one wanted to maximize the stiffness of a material to compression, $f(x)$ would have to increase as the stiffness of the corresponding configuration, $x$, increases.  In this way, eqn. \ref{rep1} attempts to increase the probability of finding the system in states with better than average values of $f(x)$: if a configuration $x$ has a value of $f(x)$ greater than the average, then the probability of finding the system near $\rho(x)$ increases, for configurations that are average in terms of $f(x)$ the probability does not change, and for those worse than average the probability decreases.  

As written, Eqn. \ref{rep1} assumes that $\rho(x|\lambda_i)$ can be independently set for every possible point  in the configuration space, in spite of the fact that $\rho(x|\lambda_i)$ is constrained to represent the physics behind the material of interest.  In actuality, $\rho(x|\lambda_i)$ can only offer a limited flexibility through the parameters $\lambda_i$.  Thus, for many problems it will not be possible to exactly satisfy eqn. \ref{rep1}, although it is possible to make a best approximation to  eqn. \ref{rep1}, given a particular physical distribution.  We achieve this by setting the changes to $\lambda_i$ such that they minimize the average error between the updates implied by eqn. \ref{rep1} and the actual changes to $\rho(x|\lambda_i) $.  Explicitly, we rewrite eqn. \ref{rep1} as $\dot{\lambda}_i\partial_{\lambda_i}\log[\rho]=f(x)-\langle f \rangle$ and select the $\dot{\lambda}_i$ that minimize the average value of the squared error, $\epsilon = \langle (\dot{\lambda}_i\partial_{\lambda_i}\log[\rho]- [f(x)-\langle f \rangle])^2 \rangle$.
The equations of motion for $\lambda_i$ that minimize $\epsilon$ at a given instant in time are easily found by setting the partial derivatives with respect to $\dot{\lambda}_i$ equal to zero.  After some manipulation (see the Supplementary Information) we find:
\begin{equation}\label{rep_final}
\dot{\lambda}_i (t) = \langle \partial_{\lambda_i} \log(\rho)\partial_{\lambda_j} \log(\rho)\rangle^{-1} \langle  [f(x) -\langle f(x) \rangle]\partial_{\lambda_j} \log(\rho) \rangle
\end{equation}
This equation is now a closed expression for $\lambda_i$ that depends only on expectation values.  Thus, it can function as an algorithm: one can draw samples from $\rho(x|\lambda_i)$, use the samples to evaluate the right hand side of eqn. \ref{rep_final}, and then integrate the equations of motion to generate new, improved parameter settings.

Equation \ref{rep_final}, and its motivating equation \ref{rep1}, overlap with a surprising number of different fields.  For example, the matrix elements in eqn. \ref{rep_final} resemble kinetic coefficients, suggesting the interpretation that $f(x)$ generates a thermodynamic force that pushes the system to solve the design goal \cite{landau1980statistical}.  Alternatively, eqn. \ref{rep_final}, appears in the optimization and mathematics literature where it is known as natural gradient descent: it bears the interpretation of a gradient descent method that takes the steepest step such that the change in entropy stays bounded \cite{amari1998natural,wierstra2008natural,ollivier2011information}.  Indeed, the matrix in front of eqn. \ref{rep_final} is the Fisher information metric and is constraining the driving force to move in directions of small entropy change \cite{wierstra2008natural}.   This interpretation is also associated with state-of-the-art optimizers like the covariance matrix adaptation evolution strategy (CMA-ES) \cite{akimoto2010bidirectional,ollivier2011information,Hansen2003}, however here the design parameters $\lambda_i$ are treated as random variables drawn from a Gaussian distribution irrespective of the design problem, and a version of eqn. \ref{rep_final} is used to update the mean and covariance of this auxiliary distribution.  This is in contrast to our proposition that $\lambda_i$ should be treated as deterministic variables that evolve according to eqn. \ref{rep_final} and with randomness only entering at the level of material configurations.  If the task considered is changed to finding the best fit model parameters to a given set of data, then eqn. \ref{rep_final} represents the direction in parameter space that decreases the fit error most efficiently per unit of behavioral change in the model\cite{transtrum2011geometry}.  In fact, in this scenario, one can consider regularizing eqn. 2 to produce the Levenburg-Mardquart algorithm modified to account for the geometric aspects of the optimization\cite{moreSloppyTranstrum}.  Finally, one can note that the motivating equation, eqn. \ref{rep1}, is the replicator equation from game theory and evolutionary biology \cite{cressman2005stability,oechssler2002dynamic,weibull1997evolutionary}.  Thus one could also interpret the dynamics as a process of reproduction and competition in a continuous parameter space \cite{cressman2005stability,oechssler2002dynamic}, projected onto $\rho(x |\lambda)$.

Whatever the picture, eqn. \ref{rep_final}. has a number of powerful properties.  In particular, eqn 2. is invariant to any invertible reparameterization of $\lambda_i$ including rotations, dilations, and translations in parameter space. If the reward function, $f(x)$, is chosen correctly, the velocity flow is also invariant to rank preserving changes in the design goal \cite{ollivier2011information}.  Thus there will be no difference in performance between two design problems that differ by coordinate choice over $\lambda_i$ and/or a rank preserving change in the design goal (e.g. $g(x)$ vs $exp(g(x))$), provided the initial values of $\lambda_i$ are the same.  These invariances also provide stability to the algorithm: by making the search algorithm invariant to both the goal function magnitude and the parameterization, the effect of sampling errors gets bounded in a parameterization invariant way.  Thus errors from sampling parameters in eqn. 2 will not cascade, even if the matrix in eqn. \ref{rep_final} becomes ill-conditioned.  Altogether, these features greatly simplify the optimization task since, now, the designer is free from worrying about trivial choices surrounding $\lambda_i$ and the goal function.  Details on these points can be found in the Supplementary Information.

For the task of optimizing materials, we stress one further property: by using eqn. \ref{rep_final} optimization takes place in configuration space, rather than in an auxiliary space introduced to define an optimizer.  As we will show, this gives a unique advantage in applying eqn. \ref{rep_final} to materials design: more information is used by the optimizer when updating parameters, without incurring an increase in computational cost.  The result is often a more reliable and efficient optimizer. Perhaps best of all, this optimizer is constructed by straightforwardly applying the formalism encoded in eqn. \ref{rep_final}  to the relevant statical model.  For example, when $\rho(x|\lambda_i)$ is given by the canonical ensemble, $\rho(x|\lambda_i)\propto\exp[-\lambda_i h_i(x)]$, the optimizer follows immediately from  eqn. \ref{rep_final} as $\dot{\lambda}_i = -Cov[h_i(x), h_j(x)]^{-1} Cov[h_j(x), f(x)]$.   Ultimately, eqn. \ref{rep_final} makes the transition from describing a material to designing a material in just one step.  

\section{Comparison Against Black-box Optimizers}

To demonstrate these strengths, we test our method against standard approaches that feed simulation parameters into a model by way of a black-box optimizer.  As our black-box optimizers of choice, we compare against adaptive simulated annealing (ASA) \cite{ingber2012adaptive} and the CMA-ES\cite{Hansen2003}.  In each test, we allow the optimizers a fixed budget of material simulations, since this is the dominant computational cost in a real-world materials design problem, and each simulation requires a fixed amount of computational power.   Thus in our comparisons, computational cost and number of simulations are equivalent and an efficient optimizer is one that solves a design problem simulating as few candidate materials as possible.  Implementation details for each of these problems are provided in the Supplementary Information.

As a first example, we task these two black-box algorithms and our new approach with designing a square-lattice Ising model to maximize the magnitude of its magnetization.   To do so, each method is allowed to vary the coupling constants that define the energy of spin alignments in the horizontal ($J_x$) and vertical ($J_y$) directions. To implement the black-box methods, we allow each optimizer to guess a set of coupling constants and evaluate the quality of that guess by computing the average magnetization.  We find that, without additional information, both ASA and the CMA-ES struggle when searching entirely in the zero magnetization phase.  This is an obvious consequence of the fact that the optimizer sees no variation in the quality for each parameter setting.  Consequently, it receives no guidance about how to update its parameter guesses and can at best walk randomly until finding the phase boundary (figure 1a\&b) 

\begin{figure}
\begin{center}
   \includegraphics[width=1.0\textwidth]{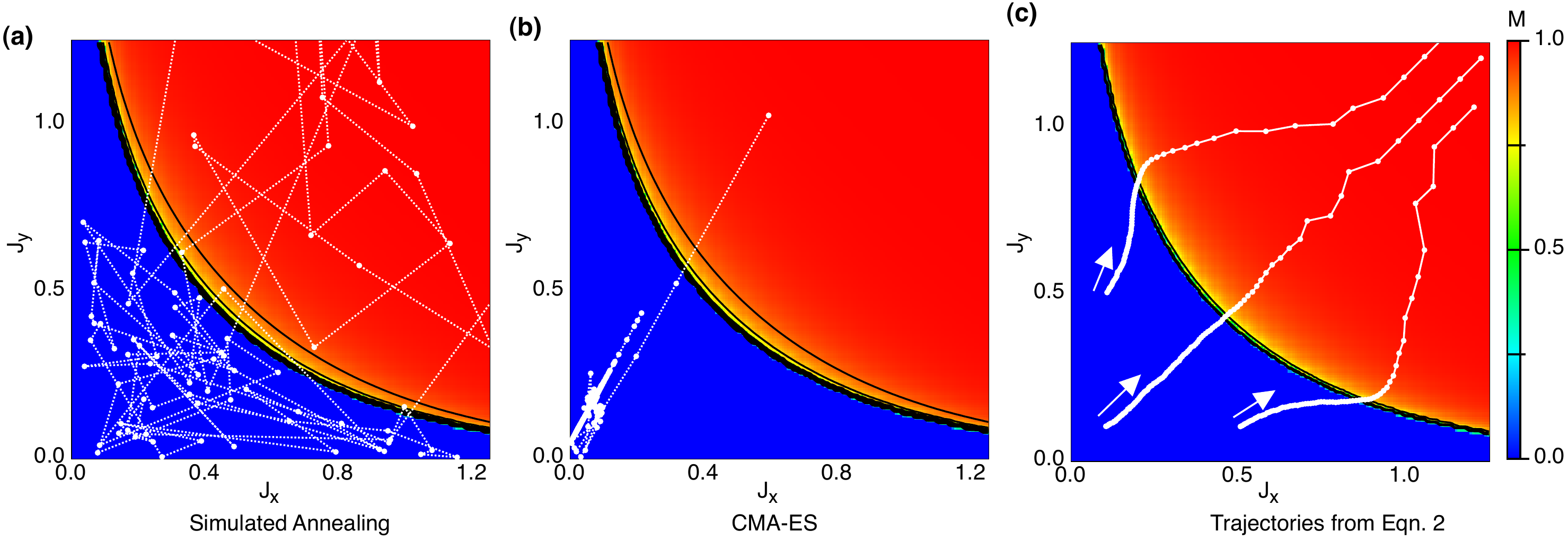}
\caption{\textbf{Phase transitions and flat landscapes.}  If a materials design problem features a phase transition as part of the search space, black-box optimizers can struggle or fail due to inefficient use of simulation data.  For example, if optimizing the magnetization, $M$, of a 2D Ising model by changing the coupling constants along the x and y directions ($J_x/kT$ and $J_y/kT$ respectively) black-box methods like simulated annealing (a) and the CMA-ES (b) walk randomly if initialized in the zero magnetization state.  Our new method (c) interprets fluctuations in solution quality against model components as encoded by eqn. 2 and thus can navigate to the magnetized state regardless of initialization.}
\label{ising}
\end{center}
\end{figure}

By contrast, the updates encoded in eqn. \ref{rep_final} navigate a path that links one phase to the other: figure \ref{ising}c shows the flow field generated by eqn. \ref{rep_final} upon the space of coupling constants $J_x/kT$ and $J_y/kT$.  This field was generated by taking $\rho(x|\lambda)$ to be the canonical ensemble with the Ising Hamiltonian, $H=-J_x \sum_{[ij]_x} s_i s_j  -J_y \sum_{[ij]_y}s_i s_j$, where $[ij]_x$ denotes summing over nearest neighbors along the x direction, likewise for $[ij]_y$, and $s_i$ denote the spin variables.  Given this statistical model, the control parameters $\lambda_i$ for optimization become $\lambda_x=J_x/kT$ and $\lambda_y=J_y/kT$.  Finally, the quality function $f(s)$ is set to reward states with higher magnetizations (see supporting material for details).  If, for shorthand, we call the individual energy components by $h_x = \sum_{[ij]_x} s_i s_j $ and $h_y= \sum_{[ij]_y} s_i s_j $, then eqn. 2 gives the velocity field $\dot{\lambda}_x = \frac{1}{|C|} (Cov[h_y, h_y] Cov[h_x, f] -Cov[h_x, h_y] Cov[h_y, f])$ where $|C|= Cov[h_x, h_x]Cov[h_y,h_y] - Cov[h_x,h_y]^2$.   A similar equation holds for $\dot{\lambda}_y$ but with the variables $x$ and $y$ appropriately interchanged.

In this form, it is clear that our method will optimize so long as there is covariation between the quality function, $f$, and the energy components, $h_i$.  Since magnetization and energy are correlated, even if the average magnetization is zero, eqn. \ref{rep_final} can purposefully optimize even when operating in regions of parameter space where the black-box methods fail.  The difference between these approaches lies in the fact that black box methods are trying to solve a problem defined over the space of $\lambda_i$, while our new approach is tasked to solve a problem defined on the space of configurations, $x$, via eqn. \ref{rep1}.  For instance the CMA-ES generates multiple guesses of parameters from a Gaussian distributed over the space of all possible $\lambda$ and ASA samples by assigning an energy value to each choice of $\lambda$.   In short, these methods only associate one piece of information, the quality function, to full ensembles of configurations defined by each choice of parameters.  This is in contrast to our new approach which tries to solve the problem of reproducing configurations, $x$, that are better than average.  Consequently, eqn. \ref{rep_final} is able to use information about how fluctuations in configurations correspond to fluctuations in quality.  That is, our new approach is unique in that it can identify relationships between the control parameters $\lambda$ and systems states $x$ because it has been built to exploit the extra fact that the simulation data were generated from a known distribution, $\rho$.

 As a second example, we consider a thermalized particle trapped on a 2D substrate, defined on the $x_1-x_2$ plane and at thermal equilibrium. The substrate applies a potential to the particle, making some positions more likely than others.  We task the optimizer with trapping the particle in a specific potential well.  To do so, we give the optimizer the freedom to tune the interaction strength with the substrate, and we give it control over a linear electric field to drive the particle.  To make the problem interesting, we use a rough substrate potential: $h_s = \sum_{i} (-\cos[x_i \frac{2 \pi}{5}] + x_i^2/25)$. With the field included, the total Hamiltonian becomes $ H= h_s +v_{x_1} x_1 +v_{x_2} x_2$, where $v_{x_1}$ and $v_{x_2}$ represent the field strength in the two coordinate directions.  These two parameters plus the temperature, $kT$, form the physical effects the optimizer may tune to solve the design problem.  To simplify the form of eqn. 2, we represent these effects to the optimizer by defining $\lambda_s = 1/kT$ and absorb a factor of $kT$ into the field coupling constants so that $\rho\propto\exp[-\lambda_s h_s - \lambda_{x_1} x_1 - \lambda_{x_2} x_2]$.   In discussing the optimizers performance, however we will convert the results the original, physical variables $kT$, $v_{x_1}$ and $v_{x_2}$. 

The solution to this problem requires the design engine to tilt the potential, make the target well the global minimum, and cool the system to zero temperature to trap the particle.  For definiteness, we say that the target well is the point $(5,5)$ and we initialize the algorithm with the substrate at $1kT$ and the field parameters set to zero.

\begin{figure}
 \begin{center}
  \includegraphics[width=1\textwidth]{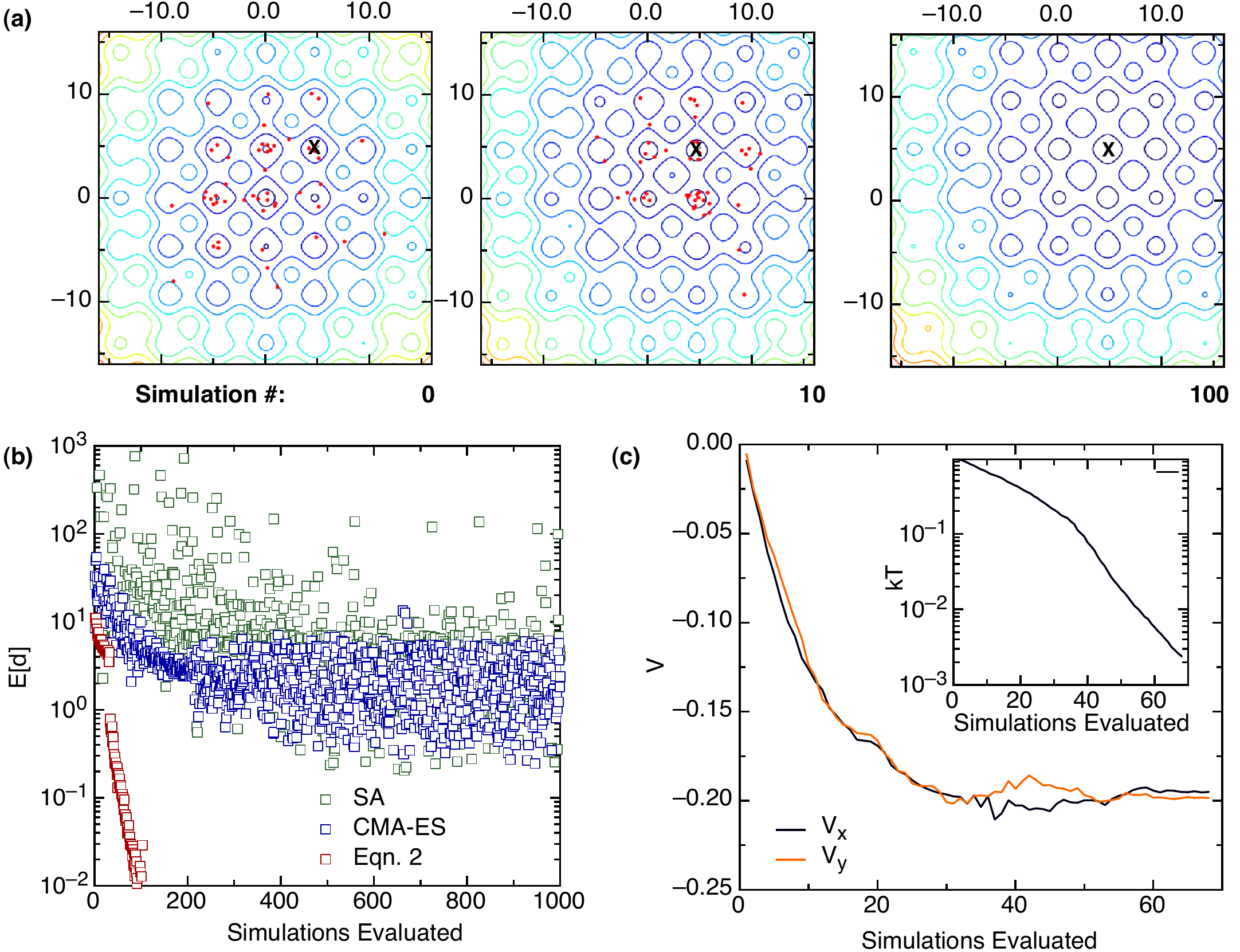}
 \caption{\textbf{Trapping a particle in a well}.  Here we treat the problem of a thermalized particle, trapped on a sinusoidal energy landscape superimposed on a quadratic background with a minima at the origin, depicted by the energy contours in (a).  The optimizer is given control over the system temperature $kT$ plus a linear field potential in the two coordinates $v_{x_1}$ and $v_{x_2}$.  Its task is to trap the particle in a specific well located off center at $(5,5)$ in the $x_1-x_2$ plane, marked by a cross in (a).   The sampled particle locations, given each choice of parameters generated, after evaluating 0, 10 and 100 iterations of eqn. 2 are plotted as red points.  The optimizer uses the field to first tilt the potential, make the target well the global minimum, and then cools the system, increasing the concentration of samples around the target.  In tasking the same problem to black-box optimizers, we find that both ASA and the CMA-ES are able to tilt the well, but never learn to cool the system.  These methods stall out producing ensemble averaged distances to the target well, $d_{ave}$, of order unity, while our method converges exponentially to a state with the particle localized at the target (b).  Comparing how the temperature and field parameters change at each iteration (c) against $d_{ave}$ in (b) shows that the exponentially convergence occurs in concert with an exponential decrease in system temperature.  Furthermore the field parameters $v_{x_i}$ are used only to align the well in the first 30 generations, and left constant during the quench.  We note that the field parameters $v_{x_1}$ and $v_{x_2}$ track one another, reflecting the fact that the optimizer is invariant to rotations in the configuration space: the optimizer moves the field along the direction associated with the greatest improvement in solution quality and is insensitive to the fact that the problem was parameterized in the arbitrary coordinates $x_1$ and $x_2$.}
\label{partLinear}
\end{center}
\end{figure}

In Fig. \ref{partLinear}a we plot the energy landscapes generated by the optimizer, as well as the points sampled during each iteration.  Indeed, the optimizer quickly learns to tilt the landscape, correctly making the target well the global minimum, and then cools the system (Fig. \ref{partLinear}b), trapping the particle deeper and tighter in the well.  By comparing the performance against black-box approaches, (Fig. \ref{partLinear}c) we observe the new approach is both faster and more reliable: it correctly tilts the well after only 35 iterations, whereas it takes a $\sim100$ simulations for the CMA-ES and ASA.  Further, neither of the black-box methods learns to completely cool the system in the allotted 1000 simulated ensembles.  

We speculate that this shortcoming is again a consequence of indirect problem representation.  We note that when $kT\approx 0.1$ the particle is almost evenly distributed between both the central well and the four nearest neighbors that surround it.  In other words, even when the substrate interaction is large compared to $kT$, the energy difference between local minima and the global optimum can remain small. Thus, for black-box methods, noise in the average particle position can play an overpowering role during parameter updates.  By contrast, eqn. \ref{rep_final} considers covariances in addition to average values and does so at the finer scale of configuration space.  Strictly using the average quality function value neglects this extra information about how certain types of fluctuations in configurations correlate to desired fluctuations in quality.  As with our prior example, this extra information makes our method appreciably more robust to flat regions in the search landscape and in this case yields an essentially exponential convergence to the optimized state (Fig. \ref{partLinear}b).

\section{Designing a Polymer to Fold into an Octahedron} 
The success of the two simple examples in the previous section invites more complicated design problems.  As an example, we consider a basic model for a polymer: a string of hard, colored balls interconnected by rigid rods.  The balls are weakly attractive, and the interactions strengths between each are determined by the colors.  For example, red and blue may be attracted more strongly than blue and green.  

In principle, by tuning the color interactions, it should be possible to fold the chain into specific, desired shapes. To make a concrete task, we take a chain of 6 particles and create an optimizer to fold them into an octahedron, defined by minimizing the sum of the distances to the center of mass \cite{Sloane1995}.  Note that the search space is appreciably larger than in the prior examples (dimension 10), and that simply setting all the interaction strengths to large values will not produce the optimal solution: the interactions are constrained to be truncated Lennard-Jones potentials.  By choosing the cutoff distance appropriately, the same energy can be given to the octahedral and polytetrahedral configurations for identical coupling constants.  Furthermore, entropic arguments imply that the polytetrahedron will dominate the chain configurations unless the optimizer carefully adjusts the coupling constants to take on unique values \cite{arkus}\cite{meng2010free}.

\begin{figure}
\begin{center}
   \includegraphics[width=0.65\textwidth]{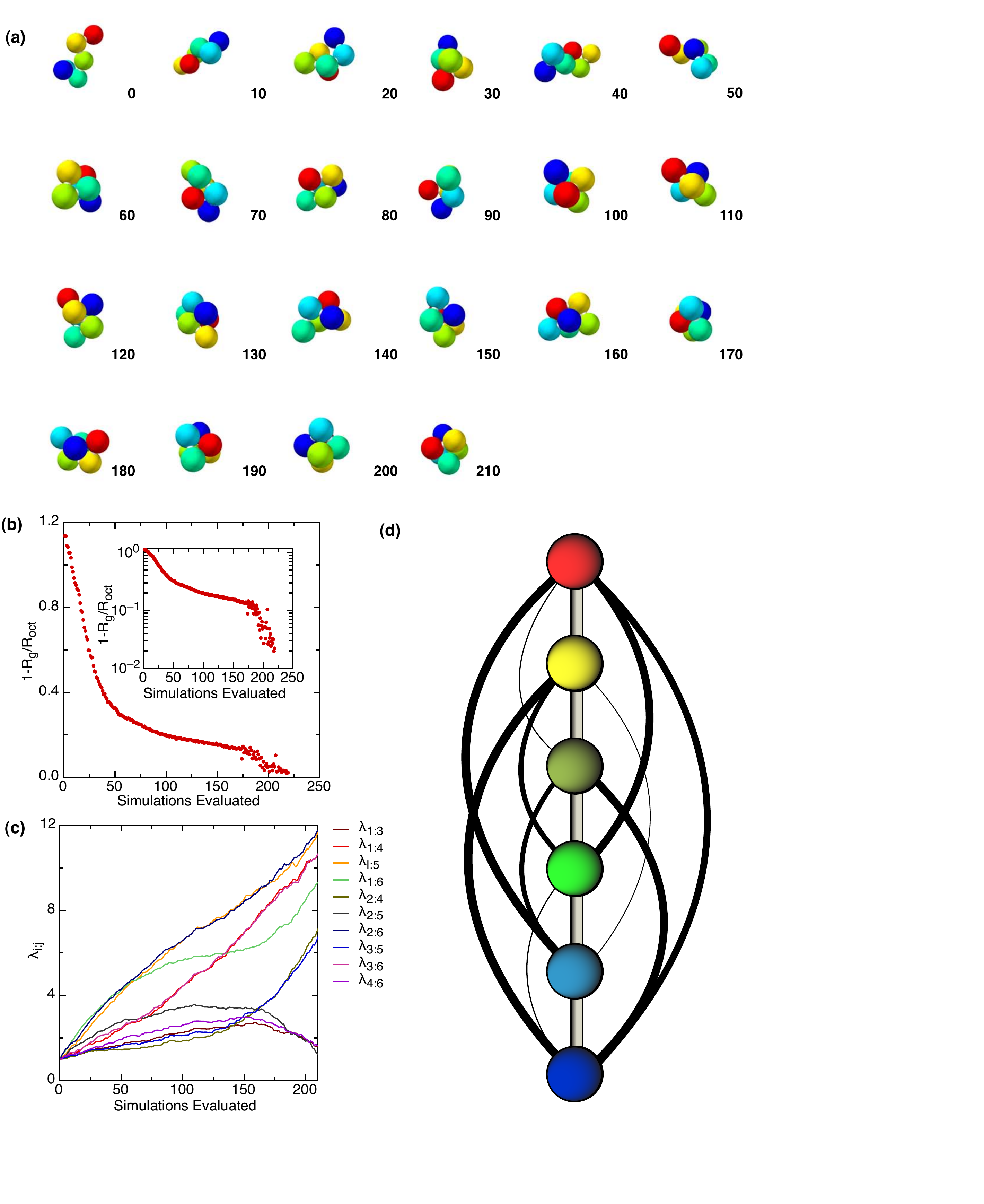}
\caption{ \label{polymer} \textbf{Folding an octahedron out of a linear chain.}  (a) Typical chain configurations that result after iterating eqn. 2.  Numbers labeling the images indicate the iteration number.  Early on, the polymer configurations are dominated by thermal energy and are random and chain-like, yet as the interaction strengths increase and differentiate, more structured objects appear, ultimately only the octahedron configuration exists (iterations 190-210).  (b) Plotting the median percent deviation between the polymer's radius of gyration $R_g$ and the radius of gyration for an octahedron $R_{oct}$ at each iteration shows that the optimizer again produces a monotonic decrease at each step, with an effectively exponential convergence in the last 30 iterations.  By the last 10 iterations, the median deviation from a perfect octahedron is roughly 1 percent.  (c) In plotting the coupling constants against iteration number, we find that the optimizer adjusts coupling constants in groups that reflect symmetry about the chain center (e.g. couplings between the first and last particles move together) and that, initially, the optimizer increases all the coupling constants in an effort to build more compact objects.  Yet this leads double tetrahedrons as the dominant chain geometry, as seen in (a).  The optimizer compensates around 100 generations by sending three of the coupling constants to zero . By plotting interactions as a network diagram (d), the choice of which three becomes clear: the remaining, active coupling constants form the contact network for an octahedron when the rigid bonds of the polymer backbone are included.  This strategy, reinterpreting the contact matrix as a guide for interaction potentials, has been developed manually as an optimal approach to self-assembly \cite{hormoz2011design}.}
\end{center}
\end{figure}

Figure \ref{polymer}a shows a typical chain configuration from each generation, while Fig. \ref{polymer}b shows the median sum  of distances to the center of mass, normalized relative to that of a perfect octahedron.   Initially, the coupling constants are set to $1kT$, and random chain configurations are typical.  However, as the optimizer drives the interaction energies to larger values, the shapes become compact and structured.  Around 200 generations, virtually every shape generated is octahedral  (Fig. \ref{polymer}a): the median deviation from that of a perfect octahedron is nearly 1 percent (Fig. \ref{polymer} b).  

By plotting the values of the interaction strengths against iteration number, we find the optimizer's solution is simple, logical and arguably optimal.  Early on, the optimizer attempts to meet the design goal by simply increasing the coupling strengths to make more compact objects (Fig. \ref{polymer}a).  However, as the coupling constants are undifferentiated, the result are predominately polytetrahedral geometries.  To compensate, the optimizer deactivates three coupling constants around 100 generations, and sends the remainder to infinity (Fig. \ref{polymer} c).  The logic behind this maneuver becomes clear by plotting the interactions as a network: the active interactions plus the polymer backbone form the contact graph of an octahedron.  This strategy, transforming the contact matrix to an interaction matrix, has been identified as an approach to programing, by hand, the optimal interaction parameters for self-assembly \cite{hormoz2011design}.  In fact, the specific problem of creating a self-assembling octahedron has been solved using a virtually identical motif \cite{zeravcic2014self}.  Altogether, these results imply exciting opportunities  for general materials design: evidently our optimizer can reproduce a well thought out approach to self assembly, and it does so automatically, requiring only a model Hamiltonian and a design goal from the user.

We note, in passing, that there is some residual in our optimizer's coupling constants.  Beyond deactivating the three interactions, some coupling constants have larger values than others.  Further, we note that the remaining seven coupling constants can be placed in four groups that respect the symmetry around the center of the chain.  For instance the binding energy between the second and last particle is almost exactly the same as the energy between the first and second to last particle (Fig \ref{polymer}c,d).  These variations show that the optimizer is responsive to the chain's influence on the likelihood of configurations, even though the chain binding the particles together is not programmed explicitly into the structure of the Hamiltonian.  In other words, these extra variations demonstrate the optimizer's capacity to learn constraints programmed implicitly in a model, and react using its explicit control parameters.

\section{Optimization of a Out of Equilibrium System}

Since eqn. \ref{rep_final} holds for any parameterized probability distribution function, it can be used to create optimization schemes beyond the canonical ensemble in the prior examples.  The only essential ingredients are a model that predicts the probability of micro-states, an engine that samples configurations from said model, and a design goal.  The optimizer can then be left to run, interacting dynamically with the adjustable parameters to produce good solutions.  As simple extensions, chemical potentials or constraints on pressure could be included as tunable parameters \cite{ferrenberg1988new}.  A new theoretical concept, termed "digital alchemy," extends statistical mechanics to account for microscale geometric parameters, such as the particle shape in a colloidal-nanoparticle assembly\cite{van2015digital}.  Thus by coupling this approach with our optimization formalism, particle geometry can be tuned to produce optimized bulk responses.  One can also note that the range of parameters to design is at the user's discretion: eqn. \ref{rep1} can be used to re-derive eqn. \ref{rep_final} assuming that some of the model parameters are not controllable by taking them to be time independent.  Indeed, for the particle in a well problem, the wavelength of corrugation was taken as a fixed parameter while the resulting optimizer was quite effective.  One can also consider optimization for global quality functions that exist over multiple a range of parameters \cite{jain2013inverse}.  For instance, suppose that one wants to optimize the density of a crystal lattice over a range of pressures and system volumes.  Our approach can be extended to this problem by defining  $\rho=\rho_0(x|\lambda, V, P)U(V, P)$ where $U(V, P)$ is a uniform distribution for $V$ and $P$ over a range of consideration and $\rho_0$ is the appropriate distribution for microstates given a fixed volume and pressure.  Eqn. 2 can then be applied to optimize in this extended parameter space to find choices of $\lambda$ that work well over a range of possible densities and pressures.  Finally, abstracting the concept, statistical models could be based on complicated computer calculations like self-consistent field theory, or just as easily, experimental measurements with the code directly measuring correlation functions in the lab and tuning physical parameters in real time. 

Moreover, non-equilibrium processes, provided they have a statistical description, are fair game.  For example, if one simulates diffusion by adding white noise to a mean drift, then the paths are distributed by a product of Gaussian distributions conditioned on the prior steps.  Clearly, the paths are statistical objects, with diffusion and drift as the distribution parameters, $\lambda$.  Thus, one can build an optimizer that tunes these control parameters using eqn. \ref{rep_final}, even if they are time dependent functions.  

As proof of this point, we return to the problem of a particle trapped on a substrate, but now simulate the particle dynamics explicitly.  The applied field and system the temperature are treated as parameterized functions of time and the optimizer is tasked with moving the particle from one fixed well to another in a given interval of time.  

\begin{figure}
\begin{center}
   \includegraphics[width=1.0\textwidth]{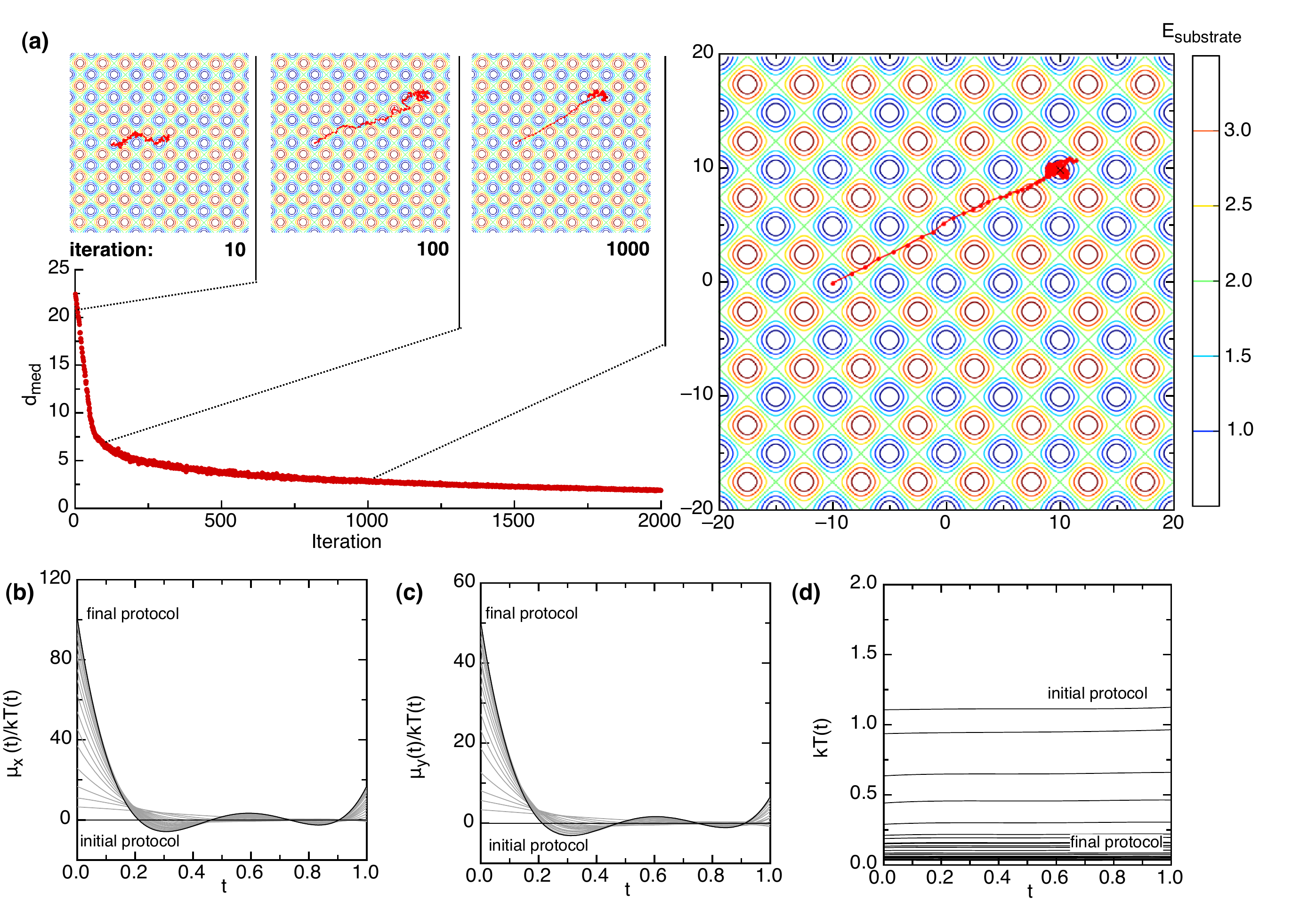}
\caption{\textbf{Optimizing a non-equilibrium process.}  By using eqn. \ref{rep_final} we can tune processing protocols for out-of-equilibrium dynamics, in this case a Brownian particle walking on a rough energy landscape controlled by a time dependent temperature, kT, and linear mean drift components, $\mu_x$ and $\mu_y$.  The optimizer has been tasked to adjust the mean applied fields and temperature to place and trap the random walker in a well located at the x-y coordinates $(10,10).$ (a) Ensemble median distance to the objective well after executing a processing protocol at each iteration of the algorithm.  Callouts show representative paths taken by the particle, and contours in the callout show lines of constant energy over the substrate potential.  The final, large image represents the ultimate protocol executed after 2000 iterations. The optimizer learns to use the driving potential effectively, in spite of the influence of the substrate, and drags the particle from the initial location to the target well. Every protocol attempted at 10 iteration intervals is illustrated in figures (b) (c) and (d), where the temperature, $kT$ (d) as well as the applied fields in the x-direction and y-direction normalized by temperature, $\mu_x /kT$ (b) and  $\mu_y /kT$ (c), are plotted against time.  Note that because the optimizer lowers $kT$ monotonically, divide the fields by it sorts them in order of their development.  At $t=0$ the particle is released from its initial position at $(-10,0)$ and allowed to wander and the processing protocol is executed until the simulation is stopped at $t=1$.   At each iteration, the optimizer works to monotonically decrease the temperature, while arriving at a field protocol that quickly drives the particle to the target well and then oscillates the fields to trap it there.}
\label{noneq}
\end{center}
\end{figure}

Figure \ref{noneq}a shows the median distance to the target well after executing the optimizer's processing protocol in each generation while callouts show the typical trajectories.   In the first 60 generations, the optimizer learns to transport the particle from its starting location to the target well via a large, deterministic driving force.  It then spends the remaining iterations monotonically decreasing the system temperature while developing a trapping protocol with the field.  After 2000 iterations, the optimizer seems to traps the particle by oscillating the driving force, changing its direction before the particle can transition to another well.  In effect, the optimizer learns to drag the particle to the target and trap it in place using both the temperature and the field.  In fact, after just these few 2000 iterations,  $\sim90\%$ of the points in the path fall within the target well.  When left to run longer, the optimizer  continues to improve the quality of solutions, but at the cost of becoming unphysical given our simulation protocol: because we did not limit the magnitude of the field strengths, the optimizer generates extremely large fields that move the particle to the well faster and faster.  To optimize beyond the proof of concept demonstrated here, one may have to restrict the range of parameters allowed to the optimizer or account for arbitrary velocities by increasing the number of steps in the walk.

\section{Optimization of Directed Self Assembly for Di-block Copolymers}

\begin{figure}
\begin{center}
   \includegraphics[width=1.0\textwidth]{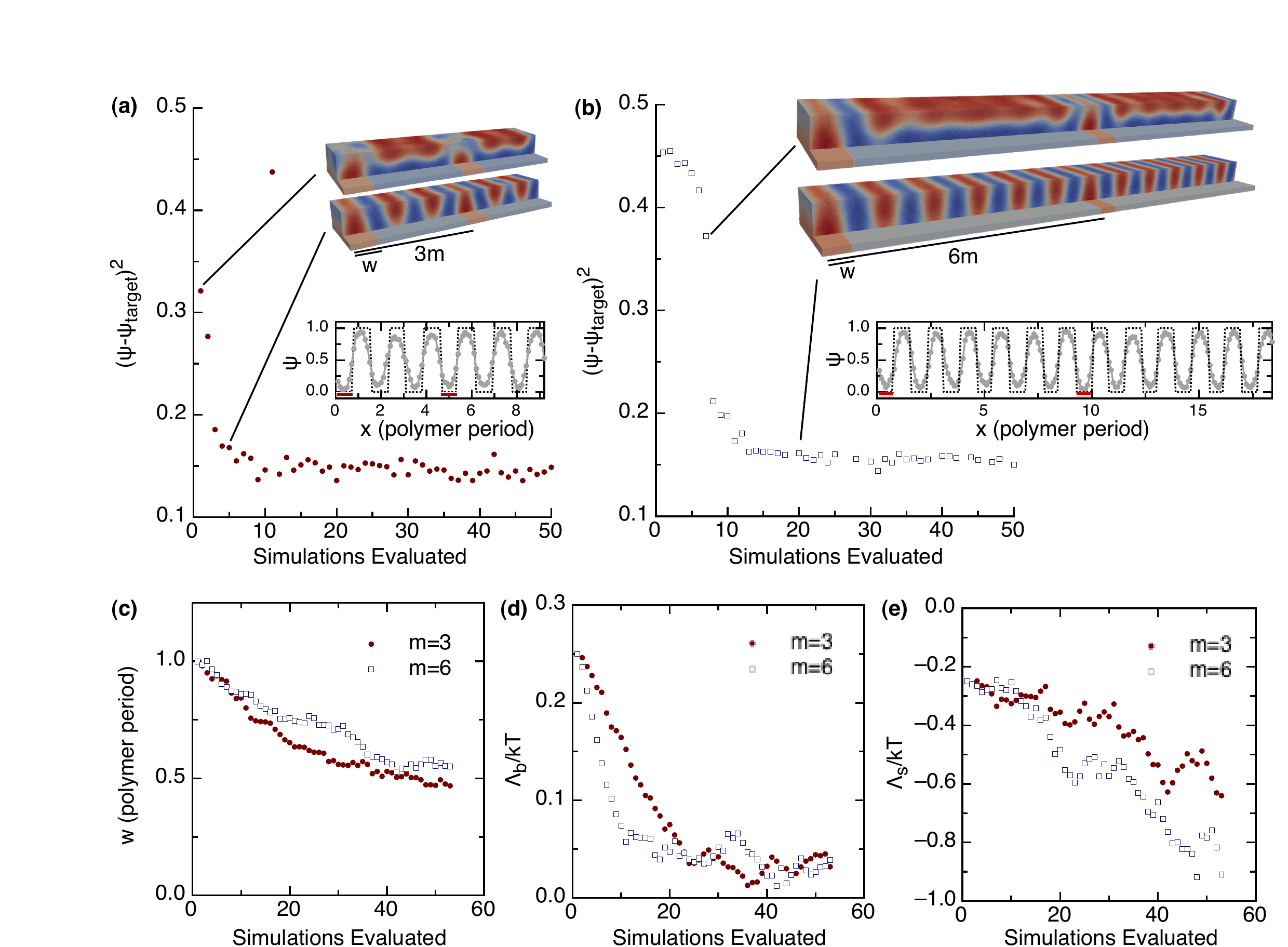}
\caption{\textbf{Optimizing directed self-assembly.} By chemically patterning stripes of width, w, that differ in their chemical affinity towards components of an AB-diblock copolymer (A depicted in blue, B depicted in red), it is possible to direct the copolymer to interpolate additional stripes up to a multiplication factor of m (a,b inset).  Here we design the width of the stripe, the strength of its attraction to the blue polymer beads $\Lambda_s$ and the attraction strength of the background substrate $\Lambda_b$ towards the red polymer beads, to match the self-assembled phase as closely as possible to the target of alternating stripes.  We quantify the success of our optimizer by comparing an order parameter $\Psi(x)= (n_a)/(n_a+n_b)$ binned along the x-axis of the box and averaged over y and z to the target stripe pattern.  By attacking this problem with eqn. \ref{rep_final}, we are able to produce optimized parameters for 3x (a) and 6x (b) density multiplication after simulating between 10 to 20 parameter choices.  Two characteristic configurations before and after optimization are plotted in the inset, separated by just a handful of iterations (a,b), yet displaying markedly different phases of the polymer.  Asymptotic configurations depicting the order parameter (solid, marked line) and the target (dashed line) (a,b, inset) show that the optimized parameters matches the desired, morphology, typically within $80\%$ or better.  In plotting the parameters generated by eqn. \ref{rep_final}, (c,d,e) we find that the interaction with the background brush is the most relevant parameter in directed self-assembly.  For both the 6x and 3x problems, the rapid convergence towards the optimized state takes place once the background strength is reduced to $\Lambda_b \approx 0.05$.  Because area of the background is significantly larger than the area of the stripe, the background must become essentially neutral before lamella can be energetically favorable.  After a weak background is established, the strip width and strength function as fine tuning parameters that facilitate defect free assembly with high reliability.}
\label{DSA}
\end{center}
\end{figure}

As a final demonstration of our approach we consider the real world problem of designing the directed self-assembly of block copolymers on a chemically patterned substrate.  This represents a task at the forefront of both materials design and sub-lithographic fabrication\cite{jain2014inverse,hannon2013inverse, hannon2013optimizing,chang2014design,Qin2013,khaira2014evolutionary,detcheverry2008monte,liu2013chemical,kim2003epitaxial,stoykovich2007directed,stoykovich2007directed}.  The goal is to lithographically pattern a substrate with a small number of chemical features such that these features promote block-copolymers to self-assemble into a desired target morphology\cite{kim2003epitaxial,stoykovich2007directed,stoykovich2007directed}.  Here, we consider a task that has been identified as a promising candidate for the manufacture of next generation semiconductor devices and high-density storage media: self-assembly of AB-diblock copolymers into an ordered striped or lamellar morphology \cite{jain2014inverse,khaira2014evolutionary,Qin2013,chang2014design},.  Specifically, we require the period of the lamella to be significantly smaller than the period of the underlying chemical pattern, as depicted in Fig. \ref{DSA}.  This requires the optimizer to adjust the interactions between the substrate and the polymer blocks so that one line of chemical patterning produces $m$ lines or periods of the block copolymer.   

We use a theoretically informed course-grain model for block copolymer simulations \cite{detcheverry2008monte}.  Polymer chains are simulated as beads that are linked together.  The system is considered at fixed temperature and volume, and thus the probability of finding a given micro-state configuration is defined in terms of an energy given by three parts.  The first is a linear spring bond energy between beads in each polymer chain.  The second is a non-bonded interaction energy that characterizes repulsion from unlike species and the material compressibility.  Details of both the bonded and non-bonded energies can be found in the literature and are summarized in the Supporting Information\cite{detcheverry2008monte}.  Both the model and parameters in it were tuned to represent a polystyrene-block-poly(methyl methacrylate) (PS-b-PMMA) diblock copolymer with a number averaged molecular weight of 22K-b-22K and a stripe period of 28nm.

The final contribution to the energy is the substrate interaction.  The substrate consists of two regions: the patterned stripes of width $w$ and the background.  Both are defined to have short-ranged effects on the polymer beads and assume the form $H_s/kT=\Lambda(\alpha)/d_s \exp[-(\frac{z}{2d_s})^2]$ where $d_s$ defines the decay length of the interaction, $z$ is the distance from the plane of the substrate, and $\Lambda(\alpha)$ is the interaction strength between the substrate and a bead of type $\alpha$.  Thus if the particle is over the guiding stripe, and of type A, $\Lambda(A) = \Lambda_s$.  If the particle is of type A and over the background region, $\Lambda(A)= \Lambda_b$.  Following \cite{khaira2014evolutionary}, we simplify our model by assuming the interactions to be antisymmetric: $\Lambda_s(A) = -\Lambda_s(B)$ and $\Lambda_b(A) = -\Lambda_b(B)$.  The design problem posed to the optimizer is to adjust the width of the strips $w$, and the two energy parameters $\Lambda_s$ and $\Lambda_b$ so the target stripe phase replicates itself $m$ times between two guiding stripes spaced by the polymer period multiplied by $m$. 

The results for $m=3$ and $m=6$ density multiplication are shown in figure \ref{DSA}.  For the $m=3$  problem, we ran the optimization 4 times varying the time-step used in integrating eqn. \ref{rep_final}.  In every instance, the optimizer not only brought the system to a state that successfully meet the design goal, but, within noise, converged to the same state each time.  The resulting, optimized parameters suggest that directed self-assembly is best achieved by setting the stripe width equal to roughly half the polymer period, $\Lambda_s\approx -1kT$ and $\Lambda_b= 0.05kT$.   All of these parameters agree with simulation results obtained by a brute-force solution to the problem, experimental verifications performed on the real polymer system \cite{liu2013chemical}, and are physically consistent with optimization results obtained for tri-block copolymer pattern multiplication\cite{khaira2014evolutionary}. These results can be explained by considering the interfacial energies in the system.  The background interaction is required to be weak since the background region has roughly equal coverage between the A and B phases, and is significantly larger in area than the size of the stripe.  Moreover, we note that the interaction strength for the stripe components is larger in the $m=6$ problem (\ref{DSA} d), which is reasonable since the larger distance between patterned regions requires stronger anchoring to guide effective assembly. 

When run at the most aggressive time-step, we were able to achieve convergence for m=3, (m=6) less than 10 (20) iterations, in spite of the fact that the material required the simulation of roughly 50,000 (100,000) polymer beads.   We stress that the performance obtained here is not a consequence of initializing the system too close to an optimal state, but rather evidence of the power behind eqn. \ref{rep_final}.  Figure \ref{DSA} shows that that indeed the initial parameters do not produce a solution to the design problem, let a lone a stripe pattern of any kind.  

We selected this particular problem because in addition to its significance for  next generation patterned media applications, it has been attempted in some variant using other materials design methods.  In fact, our initial conditions were selected to match those given to an implementation of the CMA-ES solving the same design problem but using tri-block copolymers\cite{khaira2014evolutionary}.  For that problem, the CMA-ES took roughly 50 generations to converge simulating 32 ensembles in parallel per iteration, each requiring 200,000 Monte-Carlo steps.  Our approach also used 200,000 Monte Carlo steps per iteration, but required only 10 iterations to converge.  If algorithm performance is measured in terms of the number of micro-states simulated, then solving directed self-assembly problems by way of the CMA-ES requires at least 5x as much compute power as the approach proposed here.  If it is not possible to run the ensemble simulations in parallel, our approach is roughly 130x faster than the CMA-ES, and completes a full optimization process before the CMA-ES has completed a single iteration.  Additionally, inverse  Monte-Carlo methods have been used to solve directed self-assembly problems involving the placement of guiding posts instead of stripes \cite{hannon2013optimizing}.  While there are relevant physical differences between that design problem and the one solved here, we note that the results presented for inverse Monte Carlo converge after simulating roughly 30 million micro-states.  Because this number of micro-states simulated is roughly 15 times larger than what was used here, we can speculate that our proposed methods could also be faster for such applications.

\section{Conclusions}
To the extent that the goal of materials design is a unified framework that handles a wide range of complex inverse problems, we believe the formalism introduced here represents a significant step forward.  By applying eqn. \ref{rep_final}, we can solve problems with flat search landscapes (Fig. \ref{ising}), multiple interactions types (Fig. \ref{partLinear}), incorporate constraints (Fig. \ref{polymer}), tune processing conditions (Fig. \ref{noneq}), and address application scale design and optimization tasks (Fig. \ref{DSA}). Furthermore, in all the examples presented, the end result is intuitive even though it was achieved in a complicated search landscape where other optimization schemes struggle or fail. Finally, the fact that processing conditions such as applied fields or temperature protocols and model parameters like internal interaction energies can be optimized with the very same framework presents a new direction for materials design.  Since these are the essential aspects that determine the properties of any material, the capacity to tune both simultaneously, one accounting for the other,  open the doors to a more coherent and conceptually complete design program.  

\section{Acknowledgements}
We thank Arvind Murugan, Jim Sethna, Sid Nagel, Suriyanarayanan Vaikuntanathan, and Tom Witten for many insightful discussions, and the reviewers for highly constructive suggestions..  This work was supported by the National Science Foundation through grant CBET 1334426. We acknowledge additional support through award 70NANB14H012 from the U.S. Department of Commerce, National Institute of Standards and Technology as part of the Center for Hierarchical Material Design (CHiMaD).

\section{Supplementary Information for Turning Statistical Physics Models Into Materials Design Engines}

\subsection{Equation 2. as a best approximation}
As stated in the main text, equation 2 can be derived by choosing parameter updates that minimize the average squared error $\epsilon = \langle (\dot{\lambda}_i \partial_{\lambda_i} \log[\rho] - [f(x) - \langle f\rangle])^2\rangle$.  In other words, one updates $\lambda$ so that $\dot{\lambda}$ that minimize $\epsilon$. This requires that $\partial_{\dot{\lambda}_j} \epsilon=0$ for all the $\dot{\lambda}_j$ which gives the condition
\begin{equation}
0=2 \langle \partial_{\lambda_j}\log(\rho) \partial_{\lambda_i}\log(\rho) \rangle \dot{\lambda}_i -2 \langle \partial_{\lambda_j}\log(\rho) [f(x) - \langle f \rangle]\rangle.
\end{equation}
Multiplying on the left by the pseudoinverse  for $ \langle \partial_{\lambda_j}\log(\rho) \partial_{\lambda_i}\log(\rho) \rangle$ and moving the second term to the left hand side we arrive at equation 2:
\begin{equation}\label{rep_final}
 \dot{\lambda}_i={ \langle \partial_{\lambda_i} \log(\rho)\partial_{\lambda_j} \log(\rho)\rangle }^{-1} \langle \partial_{\lambda_j} \log(\rho)[f(x) -\langle f(x)\rangle] \rangle
\end{equation}
We comment briefly on how to select an $f(x)$ that leads to a rank-invariant optimizer.  While in principle $f(x)$ can be anything that rewards solutions that conform well to the design goal, one can gain an added invariance property by designing $f(x)$ to reward good configurations based on relative rank rather than absolute value.  This makes the optimizer invariant to trivial changes in the design problem that preserve the rank of candidate solutions.  To achieve this, we took $f(x)$ to equal the probability of selecting another configuration that performs as well or worse than the current configuration $x$ when drawn randomly from the current distribution, $\rho(x|\lambda_i)$.  Explicitly, this requires

\begin{equation}
f(x) = \int dx^{\prime} \Theta(g(x^\prime)\leq g(x)) \rho(x^\prime | \lambda)
\end{equation}

where $g(x)$ is a raw quality function that determines how well solutions meet the design goal, and $\Theta(a\leq b)$ is equal to 1 whenever the inequality in the argument is satisfied and zero otherwise.  For example, in the Ising model problem $g(x)$ was taken to be the average magnetization of a given configuration, $x$.  Thus $f(x)$ rewards the configurations that have magnetizations that are likely to be larger than others drawn from the same distribution.  This rescaling indeed satisfies the requirement that any rank preserving transformation of the raw quality function $g(x)$ will not change the flow of parameters.  In other words, any rewriting of $g(x)$ that preserves the ranking of configurations will not alter the optimizer performance.  

We note in passing that by choosing this form for $f(x)$, eqn. 1 in the main text becomes the continuous space replicator equation from evolutionary game theory [29,30].  Thus we can also interpret $f(x)$ in the rank-invariant form as a fitness function that rewards the configuration at $x$  with a unit of fitness for every competitor configuration drawn from the ensemble $\rho(x)$ that performs at a lower or equal quality.

\subsection{Eqn. 2 as a projection of Eqn. 1}

Here we derive equation 2 starting from the motivating expression $\dot{\rho}(x|\lambda_i) = \rho(x|\lambda_i) [ f(x)-\langle f(x)\rangle]$ without introducing the concept of an error minimization.  We begin by noting that $\rho(x|\lambda_i)$ only has time dependance through the $\lambda_i$ terms.  Thus via the chain rule the left hand side can be expanded as
\begin{align}
\partial_{\lambda_i} \log(\rho(x|\lambda))\dot{\lambda}_i= [ f(x)-\langle f(x)\rangle]
\end{align}

To  isolate update rules for each $\lambda_i$, we project the equation onto the functions $\partial_{\lambda_j} \log(\rho)$ by multiplying both sides by  $\rho \partial_{\lambda_j} \log(\rho)  $ and integrate over configuration space:
\begin{equation}
  \langle \partial_{\lambda_j} \log(\rho)\partial_{\lambda_i} \log(\rho)\rangle \dot{\lambda}_i= \langle \partial_{\lambda_j} \log(\rho)[f(x) - \langle f(x) \rangle] \rangle
\end{equation}
The matrix on the left hand side is identified as the Fisher information matrix and can be inverted to produce 
\begin{equation}\label{rep_final}
 \dot{\lambda}_i={ \langle \partial_{\lambda_i} \log(\rho)\partial_{\lambda_j} \log(\rho)\rangle }^{-1} \langle \partial_{\lambda_j} \log(\rho)[f(x) -\langle f(x)\rangle] \rangle
\end{equation}
As desired, this form is eqn. 2 and the basis of our optimizer. 

In this derivation, we essentially multiplied  by ${ \langle \partial_{\lambda_i} \log(\rho)\partial_{\lambda_j} \log(\rho)\rangle }^{-1}  \langle \partial_j \log[\rho]|$.  Structurally, this is a projection and this gives an alternate interpretation of  eqn. 2 as the projection of eqn. 1 onto the basis functions $\partial_{\lambda_i} \log[\rho]$.
 
\subsection{Invariance Properties of Eqn. 2}
As stated in the main text, eqn. 2 hosts a range of invariance properties that yield robust performance over a large class of design problems.  One of the most important is that the velocity field generated by eqn. 2 is invariant to any reparameterizations of the design parameters $\lambda$, provided the structure of the distribution $\rho$ is left intact.  By this we mean that if there is a change of coordinates so that $\lambda_i^\prime=\lambda_i^\prime(\lambda)$ and that $\lambda_i=\lambda_i(\lambda^\prime)$ then the velocity field generated by eqn. 2 is the same in both parameterizations.  This can be shown by direct substitution:
\begin{equation}
\dot{\lambda}_i = \frac{\partial \lambda_i}{\partial \lambda_j ^\prime}\dot{\lambda}_j^\prime=\langle \frac{\partial{ \log[\rho]}}{\partial {\lambda_i}} \frac{\partial \log[\rho]}{\partial {\lambda_k} }\rangle^{-1} \langle [f-\langle f \rangle] \frac{\partial \log[\rho]}{\partial {\lambda_k} }\rangle
\end{equation}
Rewriting everything in terms of the primed variables requires the substitution $\frac{\partial  \log[\rho]}{\partial {\lambda_i}}=\frac{\partial \lambda_j^\prime}{\partial \lambda_i}\frac{\partial \log[\rho(\lambda^\prime(\lambda))]}{\partial {\lambda_j^\prime}}$ which gives us that 
\begin{align}
 \frac{\partial \lambda_i}{\partial \lambda_j ^\prime}\dot{\lambda}_j^\prime=\langle \frac{\partial \lambda_l^\prime}{\partial \lambda_i} \frac{\partial \log[\rho]}{\partial \lambda_l^\prime} \frac{\partial \lambda_m^\prime}{\partial \lambda_k} \frac{\partial \log[\rho]}{\partial \lambda_m^\prime}\rangle^{-1} \langle [f-\langle f \rangle]\frac{\partial \lambda_n^\prime}{\partial \lambda_k}\frac{\partial \log[\rho]}{\partial \lambda_n^\prime }\rangle
\\
 \frac{\partial \lambda_i}{\partial \lambda_j ^\prime}\dot{\lambda}_j^\prime=( \frac{\partial \lambda_l^\prime}{\partial \lambda_i} \langle \frac{\partial \log[\rho]}{\partial \lambda_l^\prime} \frac{\partial \log[\rho]}{\partial \lambda_m^\prime} \rangle\frac{\partial \lambda_m^\prime}{\partial \lambda_k})^{-1}\frac{\partial \lambda_n^\prime}{\partial \lambda_k} \langle [f-\langle f \rangle]\frac{\partial \log[\rho]}{\partial \lambda_n^\prime }\rangle
 \end{align}

Now we use the fact that $\delta_{ij} =\frac{ \partial{\lambda_i^\prime}}{\partial{\lambda_k}}\frac{\partial \lambda_k}{\partial \lambda^\prime_j}$ and the matrix identity $(ABA)^{-1}=A^{-1}B^{-1}A^{-1}$  to get 
\begin{equation}
\dot{\lambda}_j^\prime=\langle \frac{\partial \log[\rho]}{\partial \lambda_j^\prime} \frac{\partial \log[\rho]}{\partial \lambda_k^\prime} \rangle^{-1} \langle [f-\langle f \rangle]\frac{\partial \log[\rho]}{\partial \lambda_k^\prime }\rangle
 \end{equation}
which is the velocity equation that results when initially working from the primed coordinates.  

The key idea expressed here is that eqn. 2 assigns a velocity to each possible $\lambda_i$ in the search space that is independent of how $\lambda_i$ is parameterized.  Thus an optimizer starting at a given $\lambda_i$ position will evolve along the same trajectory as another optimizer written in a different coordinate choice.  This can provide a big advantage when compared to black box methods.  To see why, consider the fact that two different coordinate choices may lead to search spaces that are more or less corrugated in $\lambda$ space.  Since black-box methods will see the search landscapes as distinct, the performance will be better or worse depending on a trivial reparameterization of $\rho$.  By contrast, optimizers based on statistical mechanics models see no difference whatsoever, provided they are initialized to the same $\lambda$ point.  This means the designer is free to choose any coordinate system without having to worry whether the optimizer performance will degrade.

\subsection{Error propagation via Eqn. 2}
In this section we examine what effect errors in evaluating eqn. 2 can have on the algorithm performance. In particular, the matrix inverse in eqn. 2 might suggest that small errors could be amplified to produce large parameter variations.  Here we show that while small errors may indeed introduce large variations into the parameters $\lambda$, for our algorithm, this only occurs for parameters that have little effect on the simulated material properties. Specifically, we focus on implementations of eqn. 2 that use the rank based $f(x)$ defined in the previous section, integrate eqn. 2 using a small numerical time-step $\tau$, and evaluate expectation values using a Monte Carlo approximation.  To assess the fidelity of these approximations, we  compare the information lost when representing the true target distribution defined by eqn. 2 with  one constructed using the stated assumptions.  In bits, this is given by the Kullback-Leibler divergence or $K = \int dx \rho(x|\lambda_i(t))\log[\frac{\rho(x|\lambda_i(t))}{\rho(x|\lambda_i(t)+\delta \lambda_i(t))}]$, where $\delta \lambda_i(t)$ represents the error due to sampling.  If we expand to leading order in the error we find
\begin{equation}
 K= \delta \lambda_i g_{ij} \delta \lambda_j
\end{equation}
where we have used the notation $g_{ij}=\langle \partial_i \log[\rho] \partial_j \log[\rho]\rangle $.
Because we are interested in whether an ill-conditioned matrix inverse is problematic, we focus our analysis on the case where the largest contribution in error comes from computing the $\langle \partial_j \log[\rho] [f(x)-\langle f \rangle]\rangle$ terms in eqn. 2.  Thus we have that to leading order in the timestep, $\delta \lambda_i = \tau g_{ij}^{-1} \delta C_j$ where $\delta C_j$ is the error on the estimates for  $\langle \partial_j \log[\rho] [f(x)-\langle f \rangle]\rangle$.  Inserting this relationship into the equation for K and taking the expectation value over sampling realizations for $\delta C_j$ we find that to leading order  in $\tau$, 
\begin{equation}
 K=  Cov[\partial_j \log[\rho] [f(x)-\langle f \rangle], \partial_i \log[\rho] [f(x)-\langle f \rangle]]g_{ij} ^{-1}\tau^2/N
\end{equation}
where we have used the fact that, averaged over realizations, the mean squared errors for $\delta C_i$ should be given by the covariance on the estimated parameters divided by the  number of samples $N$.  We now  bound our error estimate using the fact that 
\begin{align}
 Cov[\partial_j \log[\rho] [f(x)-\langle f \rangle], \partial_i \log[\rho] [f(x)-\langle f \rangle]]g_{ij}^{-1}\leq\langle[f(x)-\langle f \rangle]^2 \partial_j \log[\rho]  \partial_i \log[\rho] \rangle g_{ij}^{-1}
 \end{align}
This equation follows because $g_{ij}^{-1}$ is a positive definite matrix and so the term $\langle  \partial_j \log[\rho] [f(x)-\langle f \rangle]\rangle g_{ij}^{-1} \langle\partial_i \log[\rho] [f(x)-\langle f \rangle]$ appearing in the covariance expression is strictly positive.  Inserting this into our expression for $K$ gives
\begin{equation}
K\leq\langle[f(x)-\langle f \rangle]^2 \partial_j \log[\rho]  \partial_i \log[\rho] \rangle g_{ij}^{-1}
 \tau^2/N
 \end{equation}

The final step is to note that the rank-based $f(x)$ used to define algorithms in this paper is bounded and the bounds are independent of $\lambda$.  Specifically, for the system described in the prior section, $0<f(x)<1$ and $\langle f \rangle = 0.5$ for all $\lambda$.  Thus, we have that $(f(x) -\langle f\rangle)^2<\frac{1}{4}$ which gives us that 
\begin{align}
K\leq\langle[f(x)-\langle f \rangle]^2 \partial_j \log[\rho]  \partial_i \log[\rho] \rangle g_{ij}^{-1}
 \frac{\tau^2}{N}\leq\frac{\tau^2}{4N} g_{ij}^{-1} g_{ij} =\frac{\tau^2 d}{4N} 
 \end{align}
where $d$ is the number of parameters $\lambda_i$ in the model.  Thus we find the surprising result that to leading order in $\tau$, the information lost by approximating eqn. 2 is bounded and that the bound is independent of both $\lambda$ and $g_{ij}$.  In other words,  to leading order in $\tau$, an approximation to eqn. 2 using the methods described in this paper will loose no more than $\frac{\tau^2 d }{4N}$ bits due to sampling noise, for any $\lambda$ selected by the optimizer.  

\subsection{Methodology for the Ising model problem}
As discussed in the main text, applying eqn. 2 to the Ising model Hamiltonian gives the optimization equation  $\dot{\lambda}_{x_l}= Cov[h_{x_l}, h_{x_m}]^{-1} Cov[h_{x_m}, f]$,  where the $h_{x_i}$ represent the energy contributions of neighbors along the horizontal and vertical directions.  To maximize the average magnetization, we set $f(x)$ equal to the probability that another configuration drawn at random from $\rho$ has a lower instantaneous magnetization, $m(x)$, than that of configuration $x$ (see Section 1, above).  Specifically, we set $f(x) = \int dx^\prime \Theta(m(x^\prime) \leq m(x)) \rho(x^\prime |\lambda)$, where $\Theta(a\leq b)$ is unity when the inequality in the argument is satisfied and zero otherwise.  With this definition all of the terms in the update equation are expressible as expectation values, and they can be evaluated by making a Monte-Carlo approximation with samples drawn from $\rho(x|\lambda)$.

To draw these samples from the Ising Hamiltonian at equilibrium, we implemented the Wolff cluster algorithm with variable coupling constants for the horizontal and vertical directions [44].   In this Markov chain Monte Carlo method, a single spin is chosen at random and flipped. The neighbors of this site are then flipped randomly based on the energy associated with the old and new configurations.  If flipped, any neighbors of the altered site are considered for manipulation, and the process iterates until no new sites are added to the cluster. For our simulations, we used a  25 by 25 grid of spin sites.  To minimize correlation between the samples, we ran a burn-in period with 1000 spin-flip cycles, logging the average number of spins flipped in each iteration.  Once this finished, we continued to iterate the spin flip algorithm and logged samples once every $q$ iterations.  We selected $q$ so that the average number of spins flipped per cycle multiplied by the number of cycles between logged samples equals the system size.  The process was repeated until 1000 configurations had been saved.  

To integrate eqn. 2 we used a modified Euler scheme with a fixed time-step of $0.5$. We suggest future work could improve this aspect of the algorithm, yet for a proof of concept this simple choice was effective.  We note that moving to higher order or variable time-step integrators will have to account for the fact that the Monte-Carlo approximation introduces error that can be larger than the truncation error associated to the integration scheme.  Thus, using more sophisticated integrators may require a careful analysis of how noise in the velocity terms impacts the optimizer trajectories.  

\subsection{Methodology for the equilibrium particle on a substrate problem}
Following the discussion in the main text, we chose our optimization parameters so that $\rho(x|\lambda_i)$ is given by a canonical ensemble of the form $\rho \propto exp[-\lambda_s h_s-\lambda_{x_1} x_1 -\lambda_{x_2} x_2]$ where the $\lambda_s$ represents control parameters for the temperature and $\lambda_{x_i}$ is the control parameter for the field contributions.  With this choice,  eqn. 2 takes the form   
\begin{equation}
\frac{d}{dt}\begin{bmatrix}\lambda_s \\ \lambda_{x_1} \\ \lambda_{x_2}\end{bmatrix} =\begin{bmatrix}Cov[h_s,h_s] & Cov[h_s,x_1] & Cov[h_s,x_2]\\ Cov[h_s,x_1] & Cov[x_1, x_1] & Cov[x_1,x_2] \\ Cov[h_s, x_2] & Cov[x_2,x_1] & Cov[x_2,x_2]\end{bmatrix}^{-1} \begin{bmatrix} Cov[h_s, f] \\ Cov[x_1,f] \\ Cov[x_2,f]\end{bmatrix}
\end{equation}

 As all of the terms in this equation are expectation values, we approximate them by averaging over samples drawn from $\rho$. Given a set of interaction strengths $\lambda_i$, we generated samples for the particle position by running a Metropolis-Hastings random walk.  The proposals were generated by adding Gaussian noise to each particle position.  For the data shown, the standard deviation of the noise was set equal to $5/ \sqrt{max[\lambda]}$. The factor of 5 corresponds to the characteristic spacing between wells in the random walk problem, and thus represents the characteristic length scale of the problem.  To decrease the correlation between samples, we discarded the first 10,000 steps in the random walk and took 50,000 more steps recording the position every 1000 cycles.  These 50 particle locations were then sent to the optimizer as samples.  

As with the Ising model example, we integrated eqn. 2 with a modified Euler scheme (time-step set to 0.5) and used a rank-based quality function for $f(x)$.  Specifically, we used the choice $f(x) = \int dx^\prime \Theta(d(x^\prime)\leq d(x)) \rho(x^\prime |\lambda)$, where $d(x)$ is the distance between $x$ and the target well and where $\Theta(a \leq b)$ is unity when the inequality in the argument is satisfied and zero otherwise.  With this definition, $f(x)$ is the probability that another configuration drawn at random from $\rho$ is further away from the target well than configuration $x$.   

\subsection{Methodology for the polymer folding problem}

By construction, we required particles to remain a fixed distance $D$ from their neighbors along the chain.  Thus we used a Metropolis-Hasting algorithm with proposal states designed to respect this condition [45, 46].  In each iteration, we selected a particle to manipulate at random.  If the particle was at the end of the chain, it was given a new position distributed uniformly at random on a unit sphere surrounding its one neighbor.   If the particle was in the interior of the chain, then all the possible new positions for the particle lived on a circle defined by the condition that the particle remains a unit distance from its two neighbors.  In this case, we assigned the interior particle a new position drawn at random and uniformly distributed on the circle. 

To deal with deep local minima that might trap the Metropolis-Hastings walk, we implemented a parallel tempering scheme [47].  We ran two other Monte-Carlo simulations in parallel with identical coupling constants, but one with a temperature twice as big and the other ten times as big.  The ensembles were allowed to attempt an exchange of configuration with the next closest temperature ensemble every 100 iterations.  The probability of exchange was set according to the standard replica-exchange algorithm [47].  With these two temperature ratios we found empirically that the probability of accepting an exchange move was roughly $0.30$ and always fell between $0.1$ and $0.5.$

We iterated this process with a burn-in period of 10,000 steps, and then took 1,000,000 steps saving samples every 1000 cycles.  The resulting 1000 configurations were then used by the optimizer to evaluate the expectation values in eqn. 2.  For our quality function, we used a rank-based reward system and defined  $f(x) = \int dx^\prime \Theta(R_g(x^\prime)\leq R_g(x)) \rho(x^\prime |\lambda)$, where $R_g(x)$ is radius of gyration for the polymer configuration $x$ and where $\Theta(a \leq b)$ is unity when the inequality in the argument is satisfied and zero otherwise.  With this definition, $f(x)$ is the probability that another configuration drawn at random from $\rho$ has a larger radius of gyration than the configuration $x$.   With all these parameters in place, eqn. 2 was integrated with a modified Euler scheme using a time-step of 0.25.  

We checked that simply setting all the energy parameters to large, identical values will not produce an octahedron.  This comes about by tuning a Lennard-Jones potential $V=4((\frac{a}{r})^{12}-(\frac{a}{r})^6)$ with a cutoff distance, $r_c$, such that $V=0$ for any $r\ge r_c$.  We shift the raw Lennard-Jones potential by its value at $r_c$ to keep the potential continuous. 

The specific value for $r_c$ was picked by ensuring that, in the octahedron geometry, the only contribution to the particle energy would come from contacting particles.  This was achieved by setting the cutoff parameter to be $r_c = \sqrt{2} D$.  Since the polytetrahedron geometry has the exact same number of particle contacts (12 contacts) and no two points closer than $r_c$, these two geometries will have exactly the same energy given the short ranged potential.

\subsection{Methodology for the non-equilibrium particle on a substrate}
Here we show how to derive an optimizer that works on an out-of-equilibrium problem.  This particular optimizer treats the problem of a particle walking randomly on a substrate and takes the random path traveled by the particle as its  configuration space.  To model the process, we discretize the path into $N$ steps each spaced in time by a small interval $\tau$.  We pick $N$ and $\tau$ so that $N\tau=1$ and set up the problem so that the random walk takes place over a time interval $0<t<1$. The path is generated by adding white noise with variance of $\tau kT_i$ and a mean drift of $\vec{v}_i$ to update the particle position $\vec{x}_i$ at each time step $i$.  Convolving these random additions gives the probability for a full path:
\begin{equation}
\rho\propto \exp[ -\sum_{i=0}^{N}\frac{1}{2\tau kT_i}(\vec{x}_{i+1} -(\vec{x}_i+\tau v(\vec{x}_i, i)))^2].
\end{equation}
By expanding the square and absorbing all the terms independent of $x_i$ into the proportionality constant we get 
\begin{equation}
\rho\propto \exp[ \sum_{i=0}^{N}-\frac{1}{2\tau kT_i}((\vec{x}_{i+1} -\vec{x}_i)^2-2(\vec{x}_{i+1} -\vec{x}_i)\vec{v}(\vec{x}_i,i) \tau)].
\end{equation}
For our particular problem, the mean velocity term $\vec{v}(\vec{x}_i, i)$ is made up of a contribution from the substrate, which depends on the particle position $-\vec{\nabla} U(\vec{x}_i)$, where $U(\vec{x}_i)=-\cos(\frac{2 \pi}{5}x)-\cos(\frac{2 \pi}{5}y)$ is the substrate potential, and a contribution from the field $\vec{\mu}_i$ that provides a directional bias and depends only on the time index $i$.  Writing these two parts explicitly gives us
\begin{equation}
\rho\propto \exp[ -\sum_{i=0}^{N}\frac{1}{2\tau kT_i}((\vec{x}_{i+1} -\vec{x}_i)^2-2(\vec{x}_{i+1} -\vec{x}_i)(\vec{\mu}_i-\vec{\nabla} U(\vec{x}_i)) \tau))].
\end{equation}
At this stage, we need to assume a particular form for how the control parameters $kT_i$ and $\vec{\mu}_i$ depend on time. One choice is to expand the parameters using a set of orthogonal basis functions.  We used the Chebyshev polynomials [48] of the first kind, $T_n(t)$.  Specifically, we define $\frac{1}{kT_i}= \sum_{n=0}^{4} (T_n(i \tau-1)+1) a_n$ for the temperature variables and $\frac{\vec{\mu}_i}{kT_i} = \sum_{n=0}^{4} T_n(i \tau-1) \vec{b}_n$ for the fields.  Note the vector notation signifies that there are two sets of $b_n$, one for the field contributions in the x-direction and one for the field component in the y-direction.  Using this choice of basis functions, we can rewrite the probability function for the path $\rho$ as 
\begin{equation}
\rho\propto \exp[  \sum_{n=0}^{4} a_n\Phi_n[x]+\vec{b}_n \vec{\Psi}_n[x]],
\end{equation}
where 
\begin{equation}
\Phi_n[x] = \sum_{i=0}^N (1+T_n(i \tau -1)) (\nabla U(x_i)-\frac{1}{2\tau}(\vec{x}_{i+1}-\vec{x}_i)^2)
\end{equation}
and
\begin{equation}
\vec{\Psi}_n[x] = \sum_{i=0}^N T_n(i \tau -1) (\vec{x}_{i+1}-\vec{x}_i).
\end{equation}
In this form, the distribution is in the same family as all the prior examples.  As far as the optimization engine is concerned, we can treat $\Phi_n$ and $\vec{\Psi}_n$ as though they were "energy" components and can produce a version of eqn. 2 using covariances between $\Phi_n$, $\vec{\Psi}_n$ and the reward function $f$:

\begin{equation}
\frac{d}{ds}\begin{bmatrix}a_n \\ \vec{b_n}\end{bmatrix} =
\begin{bmatrix}Cov [\Phi_n, \Phi_m] && Cov[\Phi_n, \vec{\Psi}_m^\dagger] \\
				Cov[\vec{\Psi}_n, \Phi_m] && Cov[\vec{\Psi_n}, \vec{\Psi}_m^\dagger] \end{bmatrix}^{-1} \begin{bmatrix} Cov[\Phi_m, f] \\ Cov[\vec{\Psi}_m, f].\end{bmatrix}
\end{equation}

Here we use the variable $s$ to denote the optimizer's time parameter so as to avoid confusion with the $t$, the time parameter for the random walk.  For the particulars of our optimizer, we again chose a modified Euler integrator with a time-step of 0.5 and calculated average values by sampling 100 paths per iteration.  We used a rank-based quality function for the $f(x)$ in eqn. 2. Specifically, we calculated the distance to the target well averaged over every point in the path $d_{target}[x]$ and then set $f(x)$ equal to the probability that another path drawn at random from the same distribution parameters has a value of $d_{target}$ greater than or equal to the given path. 

\subsection{Methodology for the Directed Self-Assembly Problem}
In this section we provide a brief overview of the key components used to model diblock copolymer directed self-assembly as well as how we implemented eqn. 2 to optimize them.   The primary focus of this section is on how one can take a generic Monte-Carlo description of a material and convert it into an optimizer via. eqn. 2.   Since the polymer model has been presented previously in the literature, we only provide a broad level description and, for details, we direct the reader to previously published work [40].  

As stated in the main text, the energy for the full system of polymer chains interacting with a chemically modified substrate is given in three parts.  The first is the bonded contribution between polymer beads in a given chain:
\begin{equation}
\frac{H_b}{kT} = \frac{3}{2} \sum_{j}^{n_p} \sum_{i}^{N-1} (r_j (i+1)-r_j(i))^2/b^2 ,
\end{equation}
where $b$ is the statistical length segment of the polymer, $N$ is the number of beads in a chain, $n_p$ is the number of polymer chains in the system, and $r_j(i)$ is the position of the i-th bead in the j-th chain.  The second contribution to our model is given by the non-bonded energy between polymer elements of different chains.  This contains two parts: one describing compressibility of the polymer melt and the other describing repulsion between different monomers:   
\begin{equation}
\frac{H_{nb}}{kT} =\frac{\sqrt{\bar{N}}}{R_e^3} \int dx[\chi N \phi_A \phi_B + (\kappa N/2) (1-\phi_A -\phi_B)^2],
\end{equation}
where $\phi_A$ and $\phi_B$ are the local densities of the A and B type beads, respectively, $R_e^2=Nb^2$ is the mean squared end-to-end distance, and the invariant degree of polymerization is given by $\sqrt{\bar{N}}= \rho_0 R_e^3 /N$, where $\rho_0$ is the average bead density.  Here  the first term characterizes the repulsion between unlike monomers and is scaled by a Flory-Huggins parameter for the model ($\chi$).  The second term defines the compressibility and its model parameter $\kappa$ is inversely proportional to the material's modulus of compression. Note that the local densities $\phi_A$ and $\phi_B$ are evaluated  using a particle-to-mesh method.

As noted in the main text, the final energy contribution comes from the particle interaction with the substrate.    Each substrate has lithographically patterned stripes of width $w$ and a background material.  The two substrate types interact with the particles by way of an energy $H_s/kT=\Lambda(\alpha)/d_s \exp[-(\frac{z}{2d_s})^2]$ where $d_s$ defines the decay length of the interaction, $z$ is the distance from the plane of the substrate, and $\Lambda(\alpha)$ is the interaction strength between the substrate and a bead of type $\alpha$.  In other words, when a particle of  type A is above the patterned region of the substrate $\Lambda(A) = \Lambda_s$, while if it is over the background region, $\Lambda(A)= \Lambda_b$.  As stated in the main text, we assume that the model parameters are antisymmetric: $\Lambda_s(A) = -\Lambda_s(B)$ and $\Lambda_b(A) = -\Lambda_b(B)$.  

Parameters for the simulation model were selected to agree with experiments performed on polystyrene-block-poly(methyl methacrylate) (PS-b-PMMA) diblock copolymers with a number-averaged molecular weight of 22K-b-22K [41]. Specifically, we took $\sqrt{\bar{N}}=83$, $\kappa N =22$ and $\chi N= 17$.  The substrate interaction length was set to $d_s=0.15$, in units of the polymer length.  Each polymer chain was comprised of 32 beads, 16 of type A followed by 16 of type B.

The design goal posed in the main text is to optimize the stripe width, $w$, and the two interaction energy scales $\Lambda_b$ and $\Lambda_s$ to promote self-assembly into a striped phase.  Since micro-state configurations are distributed according to the canonical ensemble, eqn. 2 is given by 

\begin{equation}
\frac{d}{dt}\begin{bmatrix}w \\ \Lambda_b \\ \Lambda_s \end{bmatrix} =-
\begin{bmatrix}Cov[\partial_w H_s, \partial_w H_s] && Cov[\partial_w H_s, \partial_{\Lambda_b} H_s] && Cov[\partial_w H_s, \partial_{\Lambda_s}  H_s] \\ 
Cov[\partial_{\Lambda_b} H_s, \partial_w H_s] && Cov[\partial_{\Lambda_b} H_s, \partial_{\Lambda_b} H_s] && Cov[\partial_{\Lambda_b} H_s, \partial_{\Lambda_s}  H_s] \\
Cov[\partial_{\Lambda_s} H_s, \partial_w H_s] && Cov[\partial_{\Lambda_s} H_s, \partial_{\Lambda_b} H_s] && Cov[\partial_{\Lambda_s} H_s, \partial_{\Lambda_s}  H_s]\end{bmatrix}^{-1}
\begin{bmatrix} Cov[\partial_w H_s, f] \\ Cov[\partial_{\Lambda_b}H_s,f] \\Cov[ \partial_{\Lambda_s} H_s, f] \end{bmatrix} 
\end{equation}
where we have used the fact that $-\partial_{\lambda_i} \log[\rho]= \partial_{\lambda_i} H - \langle \partial_{\lambda_i} H \rangle$.  

The partial derivatives involving $\Lambda_s$ and $\Lambda_b$ are straightforward to evaluate:  
\begin{align}
\partial_{\Lambda_s}H_s= \sum_A \theta(x,y)/d_s \exp[-(\frac{z}{2d_s})^2]-\sum_B \theta(x,y)/d_s \exp[-(\frac{z}{2d_s})^2] \\
 \partial_{\Lambda_b}H_s= \sum_A (1-\theta(x,y))/d_s \exp[-(\frac{z}{2d_s})^2]-\sum_B (1-\theta(x,y))/d_s \exp[-(\frac{z}{2d_s})^2]
 \end{align}
 where $\theta(x,y)=1$ if the particle is over a stripe and zero otherwise, and the sums $\sum_A$ and $\sum_B$ are over all beads of type $A$ and $B$.  
To evaluate the partial derivatives with respect to $w$ we decided to approximate the derivatives via a central, finite difference.  Since the method already uses a particle-to-mesh approximation, we have a small length cutoff, $l$.  This gave us the approximation 
\begin{equation}
\partial_w H_s(w, \Lambda_s \Lambda_b) \approx \frac{H_s(w+l, \Lambda_s \Lambda_b)-H(w-l, \Lambda_s \Lambda_b)}{2l}
\end{equation}

To finalize the algorithm, we used a Monte-Carlo scheme detailed in [40] to draw samples for polymers distributed in a box.  The box dimensions, as measured in units of the polymer length,  were scaled to $2m\times2\times1$, where $m$ is the multiplication target for directed  self-assembly.  Stripes were automatically drawn by the algorithm at $x=0, z=0$ and $x=m, z=0$ along the y-axis.  The polymer chains were then randomly initialized in the simulation region to fill to a desired density and where then allowed to equilibrate.  In the 3x density multiplication problem, we used 50,000 polymer beads, while in the 6x problem this was doubled to roughly 100,000.  

We allowed the system a burn-in time of 20,000 Monte-Carlo steps before we began sampling.  Samples were recorded every 10,000 steps and the full system was run for 200,000 iterations.  We decided on 10,000 steps by examining the autocorrelation time for the substrate energy contributions.  We found that after roughly 5,000 Monte-Carlo steps samples were effectively independent.  We drew samples half as frequently to be sure samples remained independent even if the optimizer explored unusual regions of parameter space.

Since this system was appreciably larger than other examples in the text, some care was taken to make integration of eqn. 2 as efficient as possible.  Our protocol was to integrate eqn. 2 using a modified midpoint method, initially taking the time-step to be 0.5.  We allowed this cursory run  100 iterations.  We found that it converged in fitness after 80 cycles and had halved the initial fitness value after roughly 20 cycles.  This suggested we could dramatically increase the time-step parameter without causing the integration to become unstable.  We scaled our time-step by a factor of 8 so that the fitness would halve after just 3 iterations and the optimization would complete in 10 cycles.  Indeed, we found this was the case in both the 3x and 6x problems as noted in the main text.  We then re-ran the optimizer 3 times at these settings to find essentially identical results in each case.

\bibliographystyle{unsrt}

\newpage

\end{document}